\documentclass[sigchi-a]{acmart}
\usepackage{booktabs} 
\usepackage{ccicons}  

\usepackage[textsize=tiny]{todonotes}
\usepackage{multirow}
\usepackage{color, colortbl}
\definecolor{DarkGray}{gray}{0.8}
\definecolor{Gray}{gray}{0.9}

\setcopyright{none}

\acmConference[Submitted to arXiv]{arXiv}{March}{2019}

\begin{document}
\title{Integrating Artificial and Human Intelligence for Efficient Translation}

\author{Nico Herbig, Santanu Pal, Josef van Genabith, Antonio Kr\"uger}
\affiliation{%
 \institution{German Research Center for Artificial Intelligence (DFKI),\\Saarland Informatics Campus}
}
\email{{nico.herbig, santanu.pal, josef.van_genabith, krueger}@dfki.de}

%
%

\renewcommand{\shortauthors}{N. Herbig et al.}

%
%

\settopmatter{printacmref=false}

\begin{marginfigure}
    \vspace{2cm}
    \includegraphics[width=\marginparwidth]{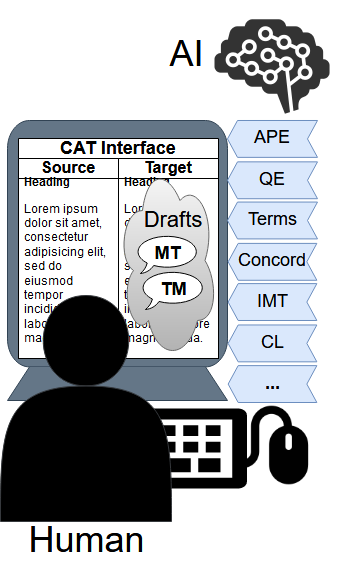}
    \caption{Overview of a human translator interacting with a variety of AI-based support tools.}
    \label{fig:titleImage}
\end{marginfigure}

%
\begin{abstract}
Current advances in machine translation increase the need for translators to switch from traditional translation to post-editing of machine-translated text, a process that saves time and improves quality. Human and artificial intelligence need to be integrated in an efficient way to leverage the advantages of both for the translation task. This paper outlines approaches at this boundary of AI and HCI and discusses open research questions to further advance the field.
\end{abstract}

%

%

%
\maketitle

\section{Introduction}
As machine translation (MT) has been making substantial improvements in recent years
, more and more professional translators are integrating the technology into their translation workflows~\cite{zaretskaya2018user}. 
The process of using a pre-translated text as a basis and improving it to the final translation is called post-editing (PE). 
It combines the advantages of both artificial intelligence (AI) and human intelligence: while the AI is good at quickly proposing draft translations of nowadays often high quality, a human with high proficiency in source and target language is needed to ensure that the meaning is identical, to analyze lexical and semantical nuances, and to understand the segment of text in a large (con)text, including the target audience, their cultural background etc. 
Apart from providing an initial translation, several AI-powered tools support this PE process and are integrated into so-called computer-aided translation (CAT) tools. 
While older research showed a strong dislike of translators towards PE~\cite{lagoudaki2009translation}, a more recent study~\cite{green2013efficacy} demonstrated that translators strongly prefer PE and argues that ``users might have dated perceptions of MT quality''. 
It has been shown that PE can not only yield productivity gains of 36\%~\cite{toral2018post}, but that it also increases the quality~\cite{green2013efficacy}.
%
This paper discusses how human and artificial intelligence can be combined for efficient language translations, which tools already exist and which open challenges remain (see Figure~\ref{fig:titleImage}).

\section{Harnessing Synergies Between AIs and Humans}

\subsection{Draft Proposal}
The PE process starts with an initial draft that is proposed by the AI and which the human uses as a basis. There are two main sources for this proposal: a machine translation (MT) and a translation memory (TM).
Simply put, TMs are large databases containing already completed human translations which are matched (using fuzzy or exact matches) against the sentence to be translated to provide a starting point for PE.
Machines can easily generate a variety of probable translations from (a combination of) MT and TM instead of only a single one; however, proposing too many and maybe even highly similar translations could overwhelm the human. The question of how to find a good set of translation suggestions to facilitate the overall translation process remains an open challenge.

\subsection{Quickly Finding and Correcting Errors}
The human's job now is to quickly find mistakes in the MT or TM outputs. So far, different support tools have been suggested for this task: 
Quality Estimation (QE) tools estimate the quality of an automatic translation without access to a reference translation~\cite{specia2009estimating}, alignments between source and MT help quickly comparing the individual parts of complex sentences and their translations~\cite{schwartz2015effects}, and color coding is used to show the similarity between input sentences and TM matches~\cite{nayek2015catalog}.
Furthermore, term bases and consistency checkers ensure consistency throughout documents, and concordance search provides ideas on how to best use words from a large corpus.
This raises the question which other tools could well support humans in finding and correcting errors within a machine-generated translation proposal. Further work should ensure that these tools do not wrongly bias the human, e.g.\ that s/he does not miss errors when the QE assumes the quality to be good.

%

\subsection{Intelligent Adaptations to Human Corrections}
Instead of only providing the best translation, MT can also be used to dynamically provide the human with alternatives for the remainder of the sentence when clicking on a word. More generally, interactive MT (IMT) tries to guess which output translation the human is going to produce given both MT input and manual changes to the input. 
Are there more appropriate visualizations than just showing the impact of changing a single word on the remainder of the sentence? Could visualizing a ``change-tree'' support the process further, or would it confuse and overload the human translator?

\subsection{Usable and Efficient Interfaces}
When comparing traditional translation to PE using all of the above support tools, one notices that the task changes considerably from mostly text production to comparing and adapting MT and TM proposals with the help of AI-powered support tools. 
Nevertheless, CAT interfaces still look very similar to interfaces used for translation from scratch and simply added those tools as additional features. 
When looking at PE interaction patterns, however, one can see that 
there is a significantly reduced amount of mouse and keyboard events~\cite{green2013efficacy}. One should therefore also explore modalities other than mouse and keyboard which could facilitate these new operations~\cite{herbig2019mmpe}.

\subsection{Avoiding Repetitive Mistakes}
Even if a human translator can efficiently translate using all the above tools and improved interfaces, s/he still does not want to correct repetitive mistakes of a MT again and again. 
A single corrected output added to the training data is very unlikely to sway a big model and there is no guarantee that the updated model is able to avoid the mistake in future. 
One approach to tackle this issue is to use Automatic Post-Editing (APE)~\cite{pal2018transformer} to incrementally adapt MT to post-edits. APE can be seen as a 2nd stage MT system, that takes the source, the first stage MT output, and the human post-edits and learns a function mapping source and MT to these post-edits. Thus, no full re-training of the large first stage MT system is required, while the corrections and individual stylistic preferences of the translator can be considered with a faster-to-train second stage model.

\subsection{Considering Human Physiology and Behavior}
Especially the PE task has the potential of inducing high cognitive load (CL) on the translator:
it involves continuous scanning of texts, including source, the incrementally evolving final translation output and possible error-prone MT output for mistakes, (sub-) strings that can be reused, text that has been translated, text that still needs to be translated, etc. 
An interesting direction for future research, with only few publications so far, is therefore to use measures proposed in HCI that allow estimating a human's CL~\cite{chen2016robust} while post-editing. For example, one could integrate measures based on e.g.\ gaze data, heart measure, or skin resistance to detect parts that are tough to read and then to automatically propose alternatives and use this feedback to better train MT systems for the future.

\subsection{Correct Bias of MT Models}
Another interesting direction of future research is to develop tools that help humans in detecting social bias in MT models. A frequently used example for such bias is the translation from the gender-neutral form ``o'' in Turkish to English, which results in translations like ``he is a doctor'' or ``she is a nurse''. While avoiding bias in AIs is being researched from the AI side, tools that help human translators to identify and correct such bias still require further research and are of the utmost importance to avoid the creation of even more biased translations that are used to train future MT systems.


\section{Conclusion}
To conclude, it will probably take a long time until machines are able to solve translation for every possible domain and every language pair without the help of humans. Therefore, researching better synergies between human and AI translators remains an important topic. The combination of AI and human intelligence leads to faster translation of higher quality.
This paper outlined a variety of existing approaches to facilitate synergies and showed future directions of research. 

%
\begin{acks}
This research was funded in part by the German research foundation (DFG) (GE 2819/2-1).
\end{acks}

%
\bibliography{sample-base}
\bibliographystyle{ACM-Reference-Format}

\end{document}